\documentstyle[aps,preprint,epsfig,amssymb]{revtex}

\newcommand{\bq}{\begin{equation}}
\newcommand{\eq}{\end{equation}}
\newcommand{\bqa}{\begin{eqnarray}}
\newcommand{\eqa}{\end{eqnarray}}
\newcommand{\nn}{\nonumber \\}

\begin{document}
\draft 
\title{Superfluid Weight vs. Superconducting Temperature based on a U(1) Slave-Boson Approach to the t-J Hamiltonian}

\author{Sung-Sik Lee and Sung-Ho Suck Salk}
\address{Department of Physics, Pohang University of Science and Technology,\\
Pohang, Kyoungbuk, Korea 790-784\\}
\date{\today}

\maketitle

\begin{abstract}
Based on an improved U(1) slave-boson approach to the t-J Hamiltonian, we investigate a relationship between the superfluid weight $n_s/m^*$(the superconducting charge carrier density/the effective mass of the charge carrier) and the superconducting temperature $T_c$.
From the present study we find a linear increase of $n_s/m^*$ with $T_c$ with the doping concentration in the underdoped region, a saturation around the optimal doping and a decrease in both $n_s/m^*$ and $T_c$ in the overdoped region.
Such a trend of the `boomerang' shaped locus in $n_s/m^*$ vs. $T_c$ with increasing doping concentration from the underdoped to the heavily overdoped region is predicted to be in complete agreement with muon-spin-relaxation measurements.
The boomerang behavior is found to occur in correlation with reduction in the spin singlet pairing(spinon pairing) order in the heavily overdoped region.
\end{abstract}
\newpage
The transverse field muon-spin-relaxation($\mu$-SR) measurements of the magnetic penetration depth $\lambda$ in high $T_c$ copper oxide superconductors reveal an universal linear increase of the superfluid weight $n_s/m^*$(the superconducting charge carrier density / the effective mass of the charge carrier) with the superconducting transition temperature $T_c$ in the underdoped region, a saturation near the optimal doping and a decrease in both $n_s/m^*$ and $T_c$ as the hole doping concentration increases from the underdoped region to the heavily overdoped region\cite{UEMURA}\cite{UEMURA_NATURE}\cite{BERNHARD}.
In the underdoped region the observed linear increase in both $n_s/m^*$ and $T_c$ scales well with the doped hole concentration $p$ while $n_s/m^*$ vs. $T_c$ does not scale with $p$ near and over the optimal doping.
One can conjecture that the observed reduction of $n_s/m^*$ with increasing hole carrier concentration in the overdoped region is attributed to either a decrease in the superconducting charge carrier density $n_s$ or to an increase in the effective mass of the charge carrier $m^*$.
In the present study we report a study of the observed boomerang behavior\cite{UEMURA_NATURE}\cite{BERNHARD} in $n_s/m^*$ vs. $T_c$ as a function of doped hole concentration by using an improved U(1) slave-boson theory which correctly allows coupling between the spin and charge degrees of freedom.
Contrary to our earlier U(1) slave-boson approach\cite{GIMM}, such introduction of both the spin and charge degrees of freedom was found to predict the arch shaped bose condensation temperature as a function of doped hole concentration $p$\cite{LEE_U1}, in agreement with the experimentally observed phase diagram in the plane of $T_c$ and $p$.
For this study we first compute the doping and temperature dependences of the superfluid weight and discuss the cause of reduction in both the superfluid weight and the superconducting temperature in the overdoped region.
In addition, a numerical comparison with the SU(2) slave-boson theory is briefly made to discuss a difference.

We write the t-J Hamiltonian,
\begin{eqnarray}
H & = & -t\sum_{<i,j>}(c_{i\sigma}^{\dagger}c_{j\sigma} + c.c.) + J\sum_{<i,j>}({\bf S}_{i} \cdot
{\bf S}_{j} - \frac{1}{4}n_{i}n_{j}) - \mu_{0} \sum_{i,\sigma} c_{i\sigma}^{\dagger}c_{i\sigma},
\label{eq:tjmodel1}
\end{eqnarray}
where ${\bf S}_{i} \cdot {\bf S}_{j} - \frac{1}{4}n_{i}n_{j} =  -\frac{1}{2} ( c_{i \downarrow}^{\dagger}c_{j \uparrow}^{\dagger}-c_{i \uparrow}^{\dagger}c_{j \downarrow}^{\dagger}) (c_{j \uparrow}c_{i \downarrow}-c_{j\downarrow} c_{i\uparrow})$.
Here ${\bf S}_{i}$ is the electron spin operator at site $i$, ${\bf S}_{i}=\frac{1}{2}c_{i\alpha}^{\dagger} \bbox{\sigma}_{\alpha \beta}c_{i\beta}$ with $\bbox{\sigma}_{\alpha \beta}$, the Pauli spin matrix element and $n_i$, the electron number operator at site $i$, $n_i=c_{i\sigma}^{\dagger} c_{i\sigma}$.
Using the single occupancy constraint and thus $c_{i \sigma}  =  b_i^\dagger f_{i \sigma}$(with $f_{i \sigma}$, spinon annihilation operator of electron spin $\sigma$ and $b_i^\dagger$, holon creation operator at site $i$), the U(1) slave-boson representation of the above t-J Hamiltonian leads to Eq.(\ref{eq:u1_sb_representation}),
\bqa
H & = & -t \sum_{<i,j>}(f_{i\sigma}^{\dagger}f_{j\sigma}b_{j}^{\dagger}b_{i} + c. c.) \nn
&& -\frac{J}{2} \sum_{<i,j>}  b_i b_j b_j^\dagger b_i^\dagger (f_{\downarrow i}^{\dagger}f_{\uparrow j}^{\dagger}-f_{\uparrow i}^ {\dagger}f_{\downarrow j}^{\dagger})(f_{\uparrow j}f_{\downarrow i}-f_{\downarrow j}f_{\uparrow i}) \nn
&& - \mu_{0} \sum_{i,\sigma} f_{i\sigma}^{\dagger}f_{i\sigma} + i\sum_{i} \lambda_{i}(f_{i\sigma}^{\dagger}f_{i\sigma}+b_{i}^{\dagger}b_{i} -1),
\label{eq:u1_sb_representation}
\eqa
where $\lambda_{i}$ is the Lagrange multiplier field to enforce the single occupancy constraint at each site.
In the SU(2) slave-boson theory, the electron operator is given by $c_{\alpha}  = \frac{1}{\sqrt{2}} h^\dagger \psi_{\alpha}$ with $\alpha=1,2$, where $\psi_1=\left( \begin{array}{c} f_1 \\ f_2^\dagger \end{array} \right)$ and  $\psi_2 = \left( \begin{array}{c} f_2 \\ -f_1^\dagger \end{array} \right)$ and $h = \left( \begin{array}{c}  b_{1} \\ b_{2} \end{array} \right)$ are respectively the doublets of spinon and holon annihilation operators in the SU(2) theory.
The SU(2) slave-boson representation of the above t-J Hamiltonian shows
\bqa
&& H   =    - \frac{t}{2} \sum_{<i,j>\sigma}  \Bigl[ (f_{\sigma i}^{\dagger}f_{\sigma j})(b_{1j}^{\dagger}b_{1i}-b_{2i}^{\dagger}b_{2j})  \nn
&& + (f_{\sigma j}^{\dagger}f_{\sigma i})(b_{1i}^{\dagger}b_{1j}-b_{2j}^{\dagger}b_{2i}) \nn
&& + (f_{2i}f_{1j}-f_{1i}f_{2j}) (b_{1j}^{\dagger}b_{2i} + b_{1i}^{\dagger}b_{2j}) \nn
&& + (f_{1j}^{\dagger}f_{2i}^{\dagger}-f_{2j}^{\dagger}f_{1i}^{\dagger}) (b_{2i}^{\dagger}b_{1j}+b_{2j}^{\dagger}b_{1i}) \Bigr] \nn
 && -  \frac{J}{2} \sum_{<i,j>} ( 1 - h_{i}^\dagger h_{i} ) ( 1 - h_{j}^\dagger h_{j} ) \times \nn
 && (f_{2i}^{\dagger}f_{1j}^{\dagger}-f_{1i}^ {\dagger}f_{2j}^{\dagger})(f_{1j}f_{2i}-f_{2j} f_{1i}) -  \mu_0 \sum_i  h_i^\dagger h_i  \nn
  && -  \sum_i  \Bigl[ i\lambda_{i}^{(1)} ( f_{1i}^{\dagger}f_{2i}^{\dagger} + b_{1i}^{\dagger}b_{2i}) + i \lambda_{i}^{(2)} ( f_{2i}f_{1i} + b_{2i}^\dagger b_{1i} ) \nn
  && + i \lambda_{i}^{(3)} ( f_{1i}^{\dagger}f_{1i} -  f_{2i} f_{2i}^{\dagger} + b_{1i}^{\dagger}b_{1i} - b_{2i}^{\dagger}b_{2i} ) \Bigr],
\label{eq:su2_sb_representation}
\eqa
where $\lambda_{i}^{(1),(2),(3)}$ are the real Lagrangian multipliers to enforce the local single occupancy constraint in the SU(2) slave-boson representation\cite{WEN}.
It is noted that in both the U(1) and SU(2) approaches above, coupling between spin and charge degrees of freedom is correctly introduced in the Heisenberg term.

After Hubbard Stratonovich transformations for the direct, exchange and pairing channels, we find the total free energy in the functional integral representation, for the U(1) slave-boson theory\cite{LEE_U1},
\bqa
&& F({\bf A}) = -\frac{1}{\beta} \ln \int D\chi D\Delta^f D\Delta^b Db Df \nn
&& e^{-(S^b({\bf A}, \chi,\Delta^f,\Delta^b,b) + S^f(\chi,\Delta^f,f)) },
\label{eq:u1_free_energy}
\eqa
where $S^b$ represents the holon action for the charge degree of freedom,
\bqa
&& S^b({\bf A},\chi,\Delta^f,\Delta^b,b) = \int_0^\beta d\tau \Big[
\sum_i b^\dagger({\bf r}_i, \tau) ( \partial_\tau - \mu^b ) b({\bf r}_i, \tau) \nn
&& + \frac{J}{2} \sum_{<i,j>} |\Delta^f_{ij}|^2 \Big[ |\Delta^b_{ij}|^2 + p^2 \Big] \nn
&& -t \sum_{<i,j>} e^{i A_{ij}} \chi_{ij} b^\dagger({\bf r}_i,\tau) b({\bf r}_j,\tau) + c.c. \nn
&& -  \frac{J}{2} \sum_{<i,j>} |\Delta^f_{ij}|^2 \Big[ \Delta^b_{ij} b^\dagger({\bf r}_i,\tau) b^\dagger({\bf r}_j,\tau) + c.c. \Big]
\label{eq:u1_holon_action}
\eqa
and $S^f$ is the spinon action for the spin degrees of freedom, 
\bqa
&& S^f(\chi,\Delta^f,f) = \int_0^\beta d\tau \Big[
\sum_i f_{\sigma}^\dagger({\bf r}_i,\tau) ( \partial_\tau - \mu^f ) f_{\sigma}({\bf r}_i,\tau) \nn
&&  + \frac{J(1- p^2)}{2} \sum_{<i,j>} \Big[ |\Delta^f_{ij}|^2 + \frac{1}{2}|\chi_{ij}|^2 \Big] \nn
&& -\frac{J}{4}(1- p )^2 \sum_{<i,j>} \chi_{ij} f_{\sigma}^\dagger({\bf r}_i,\tau) f_{\sigma}({\bf r}_j,\tau) + c.c. \nn
&& -  \frac{J}{2}(1- p )^2 \sum_{<i,j>} \Delta^f_{ij} ( f_{\downarrow}({\bf r}_i, \tau)f_{\uparrow}({\bf r}_j, \tau)  - f_{\uparrow}({\bf r}_i, \tau) f_{\downarrow}({\bf r}_j, \tau)) + c.c.  \Big].
\label{eq:u1_spinon_action}
\eqa
Here $b$ is the holon field and $f$, the spinon field.
$\chi_{ij}$, $\Delta^f_{ij}$ and $\Delta^b_{ij}$ are the Hubbard Stratonovich fields corresponding to the exchange, the spinon pairing and the holon pairing channels respectively.
After integration over the holon and spinon fields, we obtain the total free energy,
\bqa
F({\bf A}) = -\frac{1}{\beta} \ln \int D\chi D\Delta^f D\Delta^b e^{-(F^b({\bf A}, \chi, \Delta^f, \Delta^b) + F^f(\chi, \Delta^f)) },
\label{eq:free_energy_u1_2}
\eqa
where $F^b({\bf A}, \chi, \Delta^f, \Delta^b ) = -\frac{1}{\beta}\ln \int Db e^{-S^b({\bf A},\chi,\Delta^f,\Delta^b,b)}$ is the holon free energy and $F^f(\chi, \Delta^f) = -\frac{1}{\beta}\ln \int Df e^{-S^f(\chi,\Delta^f,f)}$, the spinon free energy.

The linear response of current to weak applied electromagnetic(EM) field is, in the energy-momentum space,
\bqa
j_l (\omega, {\bf q}) = -\Pi_{lm}(\omega, {\bf q})  A_m(\omega, {\bf q}),
\label{eq:london}
\eqa
with ${\bf j}$, the current and ${\bf  A}$, the EM vector potential.
Here the current response function $\Pi_{lm}(\omega, {\bf q})$ is obtained from 
\bqa
\Pi_{lm}(\omega, {\bf q}) = -\beta \left. \frac{ \partial^2}{\partial A_l(-\omega,-{\bf q}) \partial A_m(\omega,{\bf q})} F({\bf A}) \right|_{{\bf A}=0},
\label{eq:response}
\eqa
with $\beta = \frac{1}{k_B T}$, where $F({\bf A})$ is the free energy of Eq.(\ref{eq:free_energy_u1_2}) or Eq.(\ref{eq:free_energy_su2_2}).
The superfluid weight $\frac{n_s}{m^*}$ is defined as the transverse EM current response function in the static, long wave length limit\cite{DAGOTTO},
\bqa
\frac{n_s}{m^*} = \frac{1}{e^2} \lim_{ q \rightarrow 0 }  \Pi_{xx}(\omega=0,{\bf q}),
\label{eq:superfluid_weight}
\eqa
for the isotropic system.

As is shown in the second term of Eq.(\ref{eq:u1_holon_action}), the EM vector potential field $A$ is seen to modulate(is coupled with) the hopping order parameter field $\chi_{ij}$ in the U(1) slave-boson representation.
The phase fluctuations of the hopping order parameter, i.e. $\chi_{ij} = \chi e^{ i a_{ij}}$ are introduced.
For the hole pairing of d-wave symmetry we allow the d-wave spinon pairing order parameter, $ \Delta_{ji}^{f}= \pm \Delta_f$(the sign $+(-)$ is for the ${\bf ij}$ link parallel to $\hat x$ ($\hat y$)) and the s-wave holon pairing order parameter, $ \Delta_{ji}^{b}= \Delta_b$.
The EM current response function of the total system is obtained from the Ioffe-Larkin composition rule\cite{IOFFE},
\bqa
\Pi( \omega,{\bf q}) &=& \frac{\Pi^b( \omega,{\bf q}) \Pi^f( \omega, {\bf q})}{\Pi^b( \omega, {\bf q}) + \Pi^f( \omega, {\bf q})},
\label{eq:ioffe_larkin_rule}
\eqa
where $\Pi^b( \omega, {\bf q})$ is the holon current response function to the the EM field $A$ and $\Pi^f( \omega, {\bf q})$, the spinon current response function to the gauge field $a$ corresponding to the phase fluctuations of hopping order parameter.
The holon and spinon response functions are calculated from the linear response theory(see appendices 1,2).
The holon response function is evaluated to be
\bqa
\lefteqn {\Pi_{lm}^{b}( \omega, {\bf q} )  =  \frac{1}{N } \sum_{ {\bf q}_1 } 
\Big[  -2t\chi \cos q_{1l} ( u^b({\bf q_1})^2 n^b(E^b({\bf q_1})) - v^b({\bf q_1})^2 n^b(-E^b({\bf q_1}))   ) \delta_{lm} } \nn
&& + 4t^2 \chi^2 e^{i \frac{q_l - q_m}{2} } \sin ( q_{1m} + \frac{q_m}{2} ) \sin ( q_{1l} + \frac{q_l}{2} ) \times \nn
&& \Big\{
( u^b({\bf q}+{\bf q}_1)^2 u^b({\bf q}_1)^2 - u^b({\bf q}+{\bf q}_1)u^b({\bf q}_1)v^b({\bf q}+{\bf q}_1)v^b({\bf q}_1) ) \frac{ n^b(E^b({\bf q}+{\bf q}_1)) - n^b(E^b({\bf q}_1)) }{ i\omega - (E^b({\bf q}+{\bf q}_1) - E^b({\bf q}_1) ) }  \nn
&& + ( -u^b({\bf q}+{\bf q}_1)^2 v^b({\bf q}_1)^2 + u^b({\bf q}+{\bf q}_1)u^b({\bf q}_1)v^b({\bf q}+{\bf q}_1)v^b({\bf q}_1) ) \frac{ n^b(E^b({\bf q}+{\bf q}_1)) - n^b(-E^b({\bf q}_1)) }{ i\omega - (E^b({\bf q}+{\bf q}_1) + E^b({\bf q}_1) ) }  \nn
&& + ( -v^b({\bf q}+{\bf q}_1)^2 u^b({\bf q}_1)^2 + u^b({\bf q}+{\bf q}_1)u^b({\bf q}_1)v^b({\bf q}+{\bf q}_1)v^b({\bf q}_1) ) \frac{ n^b(-E^b({\bf q}+{\bf q}_1)) - n^b(E^b({\bf q}_1)) }{ i\omega + (E^b({\bf q}+{\bf q}_1) + E^b({\bf q}_1) ) }  \nn
&& + ( v^b({\bf q}+{\bf q}_1)^2 v^b({\bf q}_1)^2 - u^b({\bf q}+{\bf q}_1)u^b({\bf q}_1)v^b({\bf q}+{\bf q}_1)v^b({\bf q}_1) ) \frac{ n^b(-E^b({\bf q}+{\bf q}_1)) - n^b(-E^b({\bf q}_1)) }{ i\omega + (E^b({\bf q}+{\bf q}_1) - E^b({\bf q}_1) ) } 
\Big\}
\Big],
\label{eq:pie_b_u1_4_text}
\eqa
where $E^b({\bf q}) = \sqrt{ ( \epsilon^b({\bf q}) - \mu^b )^2 - ( J\Delta_f^2 \Delta_b \gamma_{{\bf q}})^2 }$ is the holon quasiparticle energy and $\epsilon^b({\bf q}) = -2t\chi \gamma_{{\bf q}}$, the single holon quasiparticle energy with $\gamma_{{\bf q}} = ( \cos q_x + \cos q_y )$.
$n^b(E) = \frac{1}{e^{\beta E} -1 }$ is the boson distribution function, $u^b({\bf q}) = \frac{1}{\sqrt{2}}\sqrt{\frac{\epsilon^b({\bf q})-\mu^b}{E^b({\bf q})} + 1}$ and $v^b({\bf q}) = \frac{1}{\sqrt{2}}\sqrt{\frac{\epsilon^b({\bf q})-\mu^b}{E^b({\bf q})} - 1}$.
The spinon response function is
\bqa
\lefteqn {\Pi_{lm}^{f}( \omega, {\bf q} )  =  \frac{1}{N } \sum_{ {\bf q}_1 } 
\Big[  -\frac{J}{2}(1- p )^2 \chi \cos q_{1l}  
\frac{\epsilon^f({\bf q}_1)-\mu^f}{E^f({\bf q}_1)}  ( n^f(E^f({\bf q}_1)) - n^f(-E^f({\bf q}_1)) ) \delta_{lm} } \nn
&& + ( \frac{J}{2}(1- p )^2 \chi)^2 e^{i \frac{{ q}_l - { q}_m}{2} } \sin ( {q}_{1m} + \frac{{ q}_m}{2} ) \sin ( { q}_{1l} + \frac{{ q}_l}{2} ) \times \nn
&& \Big\{
( n^f(-E^f({\bf q}+{\bf q}_1)) \frac{E^f({\bf q}+{\bf q}_1)(i\omega+E^f({\bf q}+{\bf q}_1)) + (\epsilon^f({\bf q}+{\bf q}_1)-\mu^f)(\epsilon^f({\bf q}_1)-\mu^f) + \Delta_f^{'}({\bf q}+{\bf q}_1)\Delta_f^{'}({\bf q}_1) }{ E^f({\bf q}+{\bf q}_1)[ (i\omega + E^f({\bf q}+{\bf q}_1))^2 - E^f({\bf q}_1)^2 ]}  \nn
&& + ( n^f(-E^f({\bf q}_1)) \frac{E^f({\bf q}_1)(-i\omega+E^f({\bf q}_1)) + (\epsilon^f({\bf q}+{\bf q}_1)-\mu^f)(\epsilon^f({\bf q}_1)-\mu^f) + \Delta_f^{'}({\bf q}+{\bf q}_1)\Delta_f^{'}({\bf q}_1) }{ E^f({\bf q}_1)[ (-i\omega + E^f({\bf q}_1))^2 - E^f({\bf q}+{\bf q}_1)^2 ]}  \nn
&& - ( n^f(E^f({\bf q}+{\bf q}_1)) \frac{E^f({\bf q}+{\bf q}_1)(-i\omega+E^f({\bf q}+{\bf q}_1)) + (\epsilon^f({\bf q}+{\bf q}_1)-\mu^f)(\epsilon^f({\bf q}_1)-\mu^f) + \Delta_f^{'}({\bf q}+{\bf q}_1)\Delta_f^{'}({\bf q}_1) }{ E^f({\bf q}+{\bf q}_1)[ (-i\omega + E^f({\bf q}+{\bf q}_1))^2 - E^f({\bf q}_1)^2 ]}  \nn
&& - ( n^f(E^f({\bf q}_1)) \frac{E^f({\bf q}_1)(i\omega+E^f({\bf q}_1)) + (\epsilon^f({\bf q}+{\bf q}_1)-\mu^f)(\epsilon^f({\bf q}_1)-\mu^f) + \Delta_f^{'}({\bf q}+{\bf q}_1)\Delta_f^{'}({\bf q}_1) }{ E^f({\bf q}_1)[ (i\omega + E^f({\bf q}_1))^2 - E^f({\bf q}+{\bf q}_1)^2 ]}  
\Big\} \Big],
\label{eq:pie_f_u1_4_text}
\eqa
where $E^f({\bf q}) = \sqrt{ ( \epsilon^f({\bf q}) - \mu^f )^2 + ( \Delta_f^{'}({\bf q}) )^2 }$ is the spinon quasiparticle energy, $\epsilon^f({\bf q}) = -\frac{J}{2}(1- p )^2 \chi \gamma_{{\bf q}}$, the single spinon quasiparticle energy, and  $\Delta_f^{'}({\bf q}) = J(1- p )^2 \Delta_f \varphi_{{\bf q}}$, the spinon pairing gap with $\varphi_{{\bf q}} = ( \cos q_x - \cos q_y )$.
$n^f(E) = \frac{1}{e^{\beta E} +1 }$ is the fermion distribution function.
For the U(1) theory the superfluid weight is obtained from Eq.(\ref{eq:superfluid_weight}) with the use of (\ref{eq:ioffe_larkin_rule}).
In the present calculations of the superfluid weight $J/t=0.2$ is chosen.
For other choice of $J/t$ we find that there exist no qualitative differences in the behavior of $n_s/m^*$ vs. $T_c$.

In Fig.1 the computed superfluid weight displays a linear increase in doped hole concentration(rate) $p$ in the underdoped region(the predicted optimal doping rate is $ p_o = 0.07$).
As hole doping concentration further increases, the superfluid weight first saturates and rapidly drops to $0$ in the heavily overdoped region.
This behavior in the overdoped region is attributed to a decrease in holon pairing interaction caused by the predicted diminishing trend of the spinon pairing probability amplitude(order parameter) $\Delta^f_{ij}$, as can be readily understood from the last term of Eq.(\ref{eq:u1_holon_action}).
The predicted decrease of pseudogap(spin gap) temperature with hole doping rate is in excellent agreement with observation.
In Fig.2, a ``boomerang'' shaped locus in  $\frac{n_s(T \rightarrow 0)}{m^*}$ and $T_c$ is displayed by showing a linear relationship between the two in the underdoped region and the `reflex' behavior in the overdoped region.
Although not numerically agreeable, this trend is completely consistent with muon-spin-relaxation measurements\cite{UEMURA_NATURE}\cite{BERNHARD}.
As mentioned above, the spinon pairing amplitude $\Delta^f_{ij}$ is further reduced in the overdoped region and this, in turn, causes a decrease in holon pairing interaction and thus the superconducting charge carrier density $n_s$.
This will allow the reflex behavior, that is, a decrease in both the superfluid weight and superconducting temperature. 

For the sake of introducing the low energy phase fluctuations of the order parameters $\chi_{ij}$ and $\Delta^f_{ij}$, which were not considered in the above U(1) slave-boson study, we now introduce the SU(2) slave-boson theory.
In the SU(2) theory, the EM field $A$ is found to modulate both the hopping order parameter $\chi_{ij}$ and the spinon pairing order parameter $\Delta^f_{ij}$.
The phase fluctuations of both $\chi_{ij}$ and $\Delta_{ij}^f$ are thus taken into account to assess the EM current response of the system.
We introduce the spinon pairing order parameter of d-wave symmetry, $ \Delta_{ji}^{f}= \pm \Delta_f$(the sign $+(-)$ is for the ${\bf ij}$ link parallel to $\hat x$ ($\hat y$)) and the holon pairing order parameter of s-wave symmetry, $ \Delta_{ij;\alpha \beta}^{B}= \Delta_b( \delta_{\alpha,1} \delta_{\beta,1} - \delta_{\alpha,2} \delta_{\beta,2}  )$.

In the SU(2) slave-boson theory, the response function $\Pi_{ll^{'}}^{b(A,a^1)}$ of holon isospin current to the EM field $A$ and the gauge field $a$ vanishes owing to the contribution of the $b_2$-boson in the static and long-wavelength limit(see the appendices 3,4).
Therefore the superfluid weight of the total system is given by the holon current response function only,
\bqa
\frac{n_s}{m^*} = \frac{1}{e^2} \Pi_{lm}^{b(A, A)}(\omega=0, {\bf q} = 0).
\label{eq:su2_superfluie_weight}
\eqa
Here $\Pi_{lm}^{b(A, A)}(\omega, {\bf q})$ is computed from the use of the usual linear response theory for the holon action Eq.(\ref{eq:su2_holon_action})(see Appendix 6).
In Fig.3, we show the doping dependence of the superfluid weight with the choice of $J/t=0.2$.
The superfluid weight predicted from the SU(2) theory increases faster than the U(1) case, as doped hole concentration.
This is because the gauge fields(or the phase fluctuations of the order parameters) do not screen the EM field in the SU(2) theory.
In Fig.4, the relation between $T_c$ and $\frac{n_s(T \rightarrow 0)}{m^*}$ is displayed;
at low doping, $T_c$ increases linearly with $\frac{n_s(T \rightarrow 0)}{m^*}$ as in the U(1) theory, a plateau near the optimal doping $p=0.13$) and a  reflex in the overdoped region is also observed.
There exists only a quantitative (but not qualitative) difference between the two theories.
This is because the spinon pairing amplitude $\Delta^f_{ij}$ predicted by the SU(2) theory slowly diminishes in the overdoped region compared to the U(1) case.

In summary, we investigated a relationship between the superfluid weight and the superconducting temperature, that is, $\frac{n_s}{m^*}$ vs. $T_c$ based on both the U(1) and SU(2) slave-boson approaches to the t-J Hamiltonian.
From both theories we find a qualitatively similar boomerang shape in the path of $\frac{n_s(T \rightarrow 0)}{m^*}$ vs. $T_c$ by showing a linear increase in the underdoped region and a reflex behavior in the overdoped region.
Such trend of boomerang behavior in high $T_c$ superconductors is completely consistent with $\mu$-SR experiments.
The reflex behavior predicted by both theories is attributed to the diminishing attractive hole pairing interaction caused by the markedly reduced spin pairing order $\Delta^f$, particularly in the heavily overdoped region, and thus to the reduction of both the superconducting charge carrier density $n_s$ and the superconducting temperature $T_c$ in the overdoped region.


\newpage
\centerline{FIGURE CAPTIONS}
\begin{itemize}

\item[Fig. 1 ]
The doping dependence of the superfluid weight $\frac{n_s}{m^*}$ in the U(1) slave-boson theory at $T/t=0.001$ with $J/t=0.2$. $p$ is the hole doping rate.

\item[Fig. 2 ]
The plot of $\frac{n_s(T = 0.001t )}{m^*}$ vs $T_c$ in the U(1) slave-boson theory with $J/t=0.2$.
The thick arrows point towards increasing doping rate.
$ p_o = 0.07$ is the predicted optimal doping rate.

\item[Fig. 3 ]
The doping dependence of the superfluid weight $\frac{n_s}{m^*}$ in the SU(2) slave-boson theory at $T/t=0.001$ with $J/t=0.2$. $p$ is the hole doping rate.

\item[Fig. 4 ]
The plot of $\frac{n_s(T=0.001t)}{m^*}$ vs $T_c$ in the SU(2) slave-boson theory with $J/t=0.2$.
The thick arrows point towards increasing doping rate.
$ p_o = 0.13$ is the predicted optimal doping rate.

\end{itemize}

\newpage
\begin{figure}
	\epsfxsize=9cm
	\epsfysize=5.9cm
	\epsffile{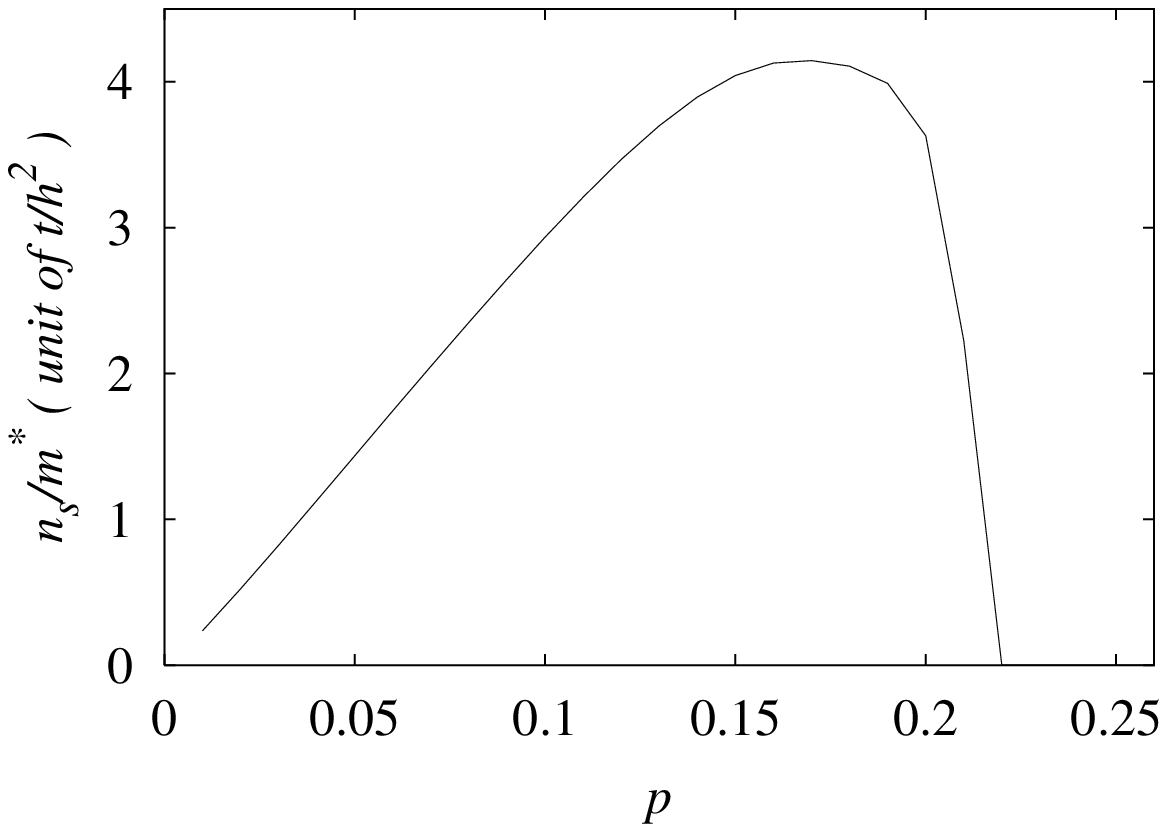}
\label{fig:1}
\caption{}
\end{figure}

\begin{figure}
	\epsfxsize=9cm
	\epsfysize=5.9cm
	\epsffile{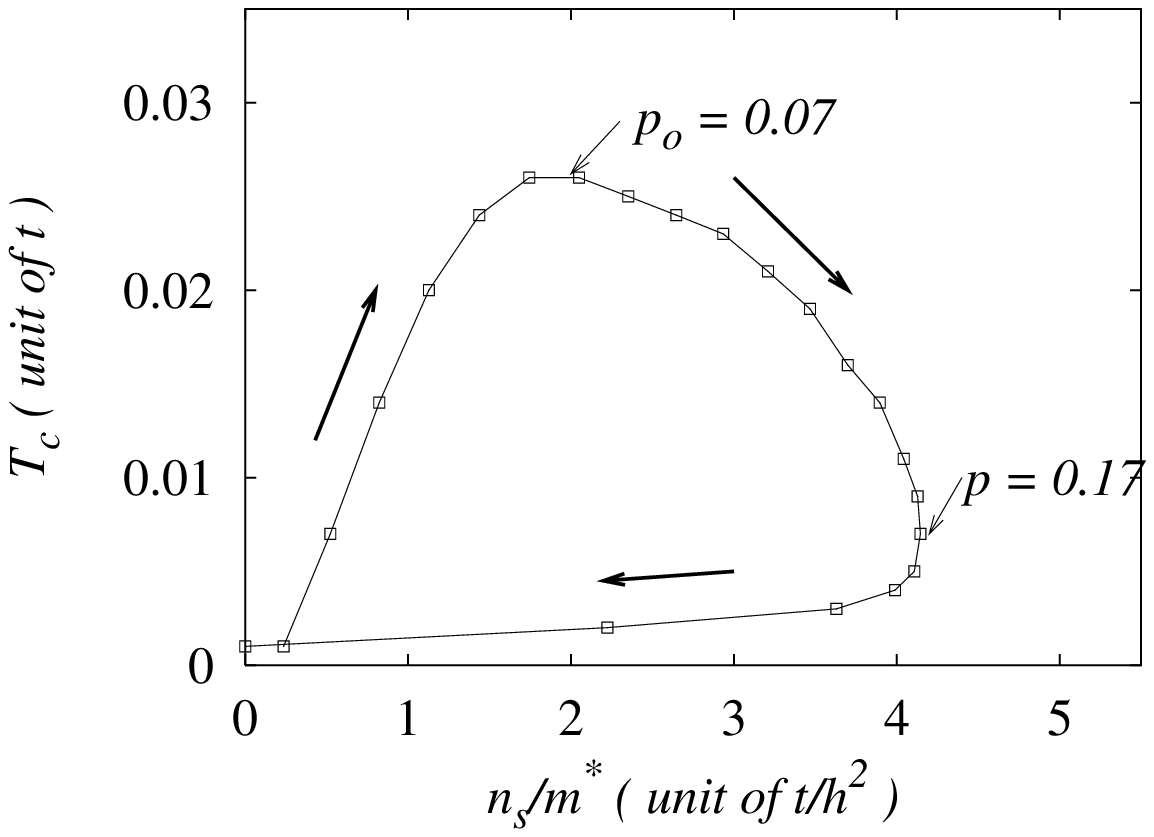}
\label{fig:2}
\caption{}
\end{figure}

\begin{figure}
	\epsfxsize=9cm
	\epsfysize=5.9cm
	\epsffile{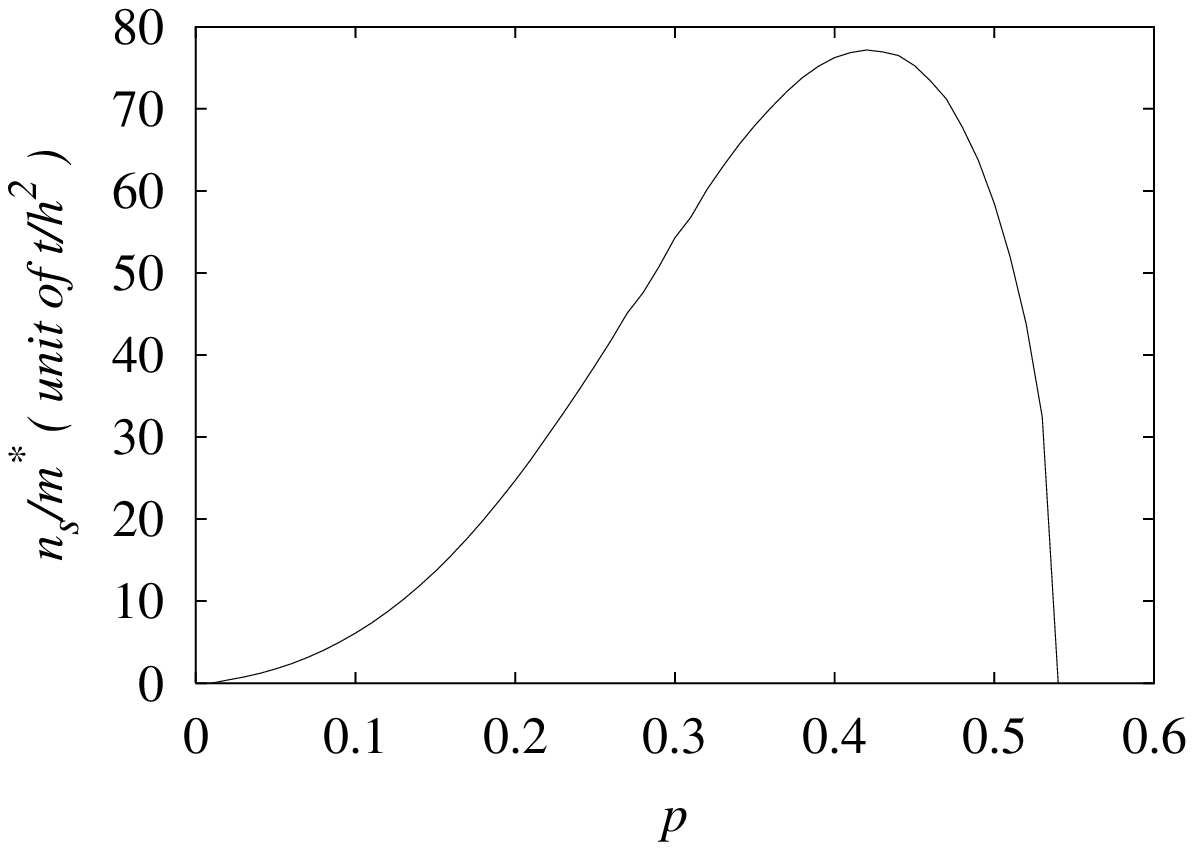}
\label{fig:3}
\caption{}
\end{figure}

\begin{figure}
	\epsfxsize=9cm
	\epsfysize=5.9cm
	\epsffile{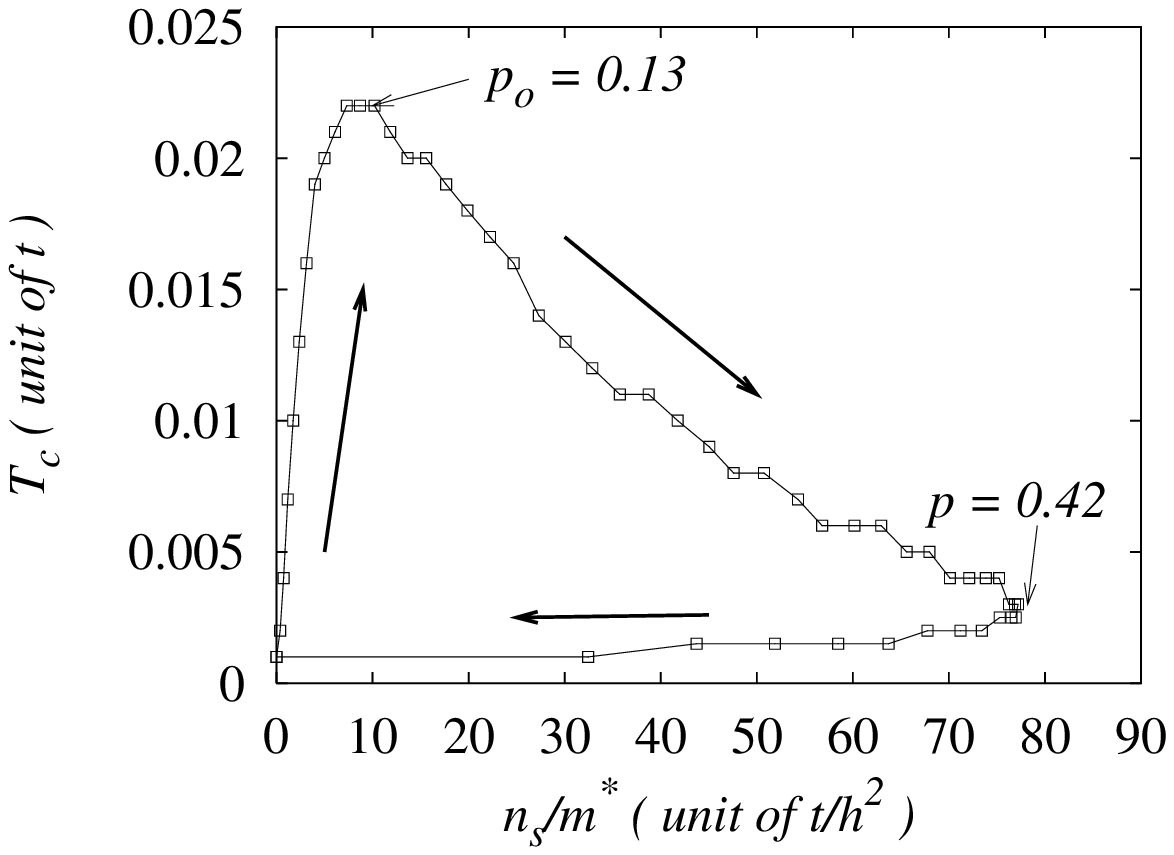}
\label{fig:4}
\caption{}
\end{figure}






\references
\bibitem{UEMURA}  Y. J.  Uemura et al, Phys. Rev. Lett. {\bf 62}, 2317 (1989); Y. J.  Uemura et al, Phys. Rev. Lett. {\bf 66}, 2665 (1991).
\bibitem{UEMURA_NATURE}  Y. J.  Uemura, A. Keren, L. P. Le, G. M. Luke, W. D. Wu, Y. Kubo, T. Manako, Y. Shimakawa, M. Subramanian, J. L. Cobb and J. T. Markert, Nature {\bf 364}, 605 (1993). 
\bibitem{BERNHARD} C. Bernhard, Ch. Niedermayer, U. Binninger, A. Hofer, Ch. Wenger, J. L. Tallon, G. V. M. Williams, E. J. Ansaldo, J. I. Budnick, C. E. Stronach, D. R. Noakes, and M. A. Blankson-Mills, Phys. Rev. B {\bf 52}, 10488 (1995); references therein.
\bibitem{GIMM} T.-H. Gimm, S.-S. Lee, S.-P. Hong and Sung-Ho Suck Salk, Phys. Rev. B {\bf 60}, 6324 (1999).
\bibitem{LEE_U1} S.-S. Lee and Sung-Ho Suck Salk, Int. J. Mod. Phys. {\bf 13}, 3455 (1999)(cond-mat/9905268). 
\bibitem{DAGOTTO} E. Dagotto, Rev. Mod. Phys. {\bf 60}, 763 (1994).
\bibitem{KOTLIAR} G. Kotliar and J. Liu, Phys. Rev. B {\bf 38}, 5142 (1988); references there-in.
\bibitem{UBBENS} a) M. U. Ubbens and P. A. Lee, Phys. Rev. B {\bf 46}, 8434 (1992); b) M. U. Ubbens and P. A. Lee, Phys. Rev. B {\bf 49}, 6853 (1994); references there-in.
\bibitem{WEN} a) X. G. Wen and P. A. Lee, Phys. Rev. Lett. {\bf 76}, 503 (1996); b) X. G. Wen and P. A. Lee, Phys. Rev. Lett. {\bf 80}, 2193 (1998); references there-in.
\bibitem{LEE_SU2} Sung-Ho Suck Salk and S.-S. Lee, Physica B, in press(cond-mat/9907226).
\bibitem{IOFFE} L. Ioffe and A. Larkin, Phys. Rev. B {\bf 39}, 8988 (1989).
\bibitem{NAGAOSA_92} N. Nagaosa and P. A. Lee, Phys. Rev. B {\bf 45}, 966 (1992).
\end{document}